# Discrete Spectrum of ULF Oscillations of the Ionosphere

A.V. Guglielmi[1], B.I. Klain[2], A.S. Potapov[3]

[1] *Schmidt Institute of Physics of the Earth, Russian Academy of Sciences, Moscow, Russia,* *guglielmi@mail.ru*

[2] *Borok Geophysical Observatory, the Branch of Schmidt Institute of Physics of the Earth, Russian Academy of Sciences, Borok, Yaroslavl Region, Russia,* *klb314@mail.ru*

[3] *Institute of Solar-Terrestrial Physics of Siberian Branch of Russian Academy of Sciences, Irkutsk, Russia,* *potapov@iszf.irk.ru*

**Abstract.** This work relates to the field of ultra-low-frequency (ULF) spectroscopy of near-Earth plasma. Our attention is focused on the Alfvén ionospheric resonator (IAR), the spectrum of which is a series of discrete lines. The aim of the study is to experimentally verify a well-defined theoretical prediction. Within the framework of an idealized resonator model, a hypothesis about the relationship between the frequencies of spectral lines has been formulated. To test the hypothesis, observations of ULF oscillations at the mid-latitude Mondy observatory were used. The result of the analysis confirmed the hypothesis that odd harmonics of the IAR discrete spectrum are observed on the Earth's surface.

## 1. Introduction

Electromagnetic oscillations of natural origin have an extraordinary variety of forms. Among them, oscillations with a dynamic spectrum in the form of a set of discrete lines are distinguished by their uniqueness. In the ultra-low-frequency (ULF) range, which extends from millihertz to several hundred hertz [Guglielmi, Pokhotelov, 1996], a typical example is magnetosonic oscillations at high harmonics of the ion gyrofrequency existing in a toroidal waveguide under the arch of the plasmasphere [Guglielmi, Klain, Potapov, 1975 ]. Another example is given by the theory of an open ion-cyclotron resonator, presumably existing in the outer radiation belt of the Earth [Guglielmi, Potapov, Russell, 2000]. The resonator spectrum has an unusual form of closely spaced lines lying above the gyrofrequency of heavy ions. Similar structures also exist in the radio range. Let us point out here the so-called "zebra-structure" of the spectrum of one of the types of solar radio emission [Zlotnik, 2010].

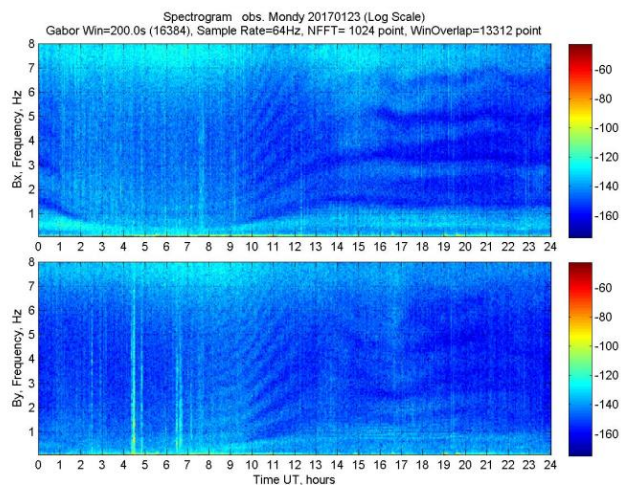

**Figure 1.** The dynamic spectrum of the electromagnetic oscillations of the ionosphere recorded at the Mondy observatory of the ISTP SB RAS on January 23, 2017.

In this paper, we will focus on the discrete spectrum of ionospheric oscillations that are observed in the Pc1 subband (0.2–5 Hz) of the total ULF oscillation range. Figure 1 gives an idea of the structure and dynamics of spectral lines. Oscillations were registered on January 23, 2017 at the Mondy mid-latitude observatory of the ISTP SB RAS (52° N, 101° E). We see a ribbed structure in the form of a set of harmonics; this is especially noticeable after 08 UT. The frequencies of the harmonics change smoothly in time, which is naturally associated with the diurnal variation of the ionospheric parameters.

ULF oscillations of the ionosphere were discovered in [Belyaev et al., 1987] and were called ionospheric Alfvén resonances (IAR). The name reflects the prevailing concept of IAR as of standing Alfven waves, which are excited in the ionosphere by external sources (lightning discharges in the atmosphere, electromagnetic noise penetrating into the ionosphere from the magnetosphere), and are also self-excited as a result of the instability of the ionospheric plasma. Extensive literature is devoted to the results of experimental and theoretical studies of IAR [Polyakov, Rapoport, 1981; Belyaev et al., 1990; Lysak, 1991; Demekhov et al., 2000; Pokhotelov et al., 2000; Yakhnin et al., 2003; Molchanov et al., 2004; Bösinger et al., 2004; Lysak, Yoshikawa, 2006; Potapov et al., 2014, 2017; Baru et al., 2016; Fedorov et al., 2016; Dovbnya, Klain, Kurazhkovskaya, 2019a, b]. A more complete bibliography can be found in the review [Potapov et al., 2021].

This work is related to the field of ULF spectroscopy of near-Earth plasma. We focused our attention on the important issue of the distribution of the IAR spectral lines. The aim of the study is to experimentally verify a well-defined theoretical prediction. The statement of the problem is presented in Section 2. Within the framework of the idealized model of the ionospheric resonator, we formulated a simple hypothesis about the relationship between the frequencies of spectral lines. The result of testing the hypothesis on the basis of IAR observations is presented in Section 3. Section 4 is devoted to a discussion of the result obtained.

## 2. Idealized resonator model

We use a simple IAR model to present the idea of this experimental study. Let us represent the ionosphere in the form of a homogeneous flat horizontal layer of dense plasma in a homogeneous external magnetic field **B**, the field lines of which are vertical. Above the ionosphere is a homogeneous half-space filled with rarefied plasma, which simulates the magnetosphere. Below the ionosphere is the atmosphere, which can be considered as a vacuum layer, under which the perfectly conducting Earth's crust is located. Let the Alfvén velocity $c_A$ in the ionosphere be much less than in the magnetosphere, so that the reflection coefficient of Alfvén waves incident from below on the upper boundary of the ionosphere can be considered indistinguishable from unity. This is a very crude model, but it qualitatively correctly reflects the general idea of the IAR structure.

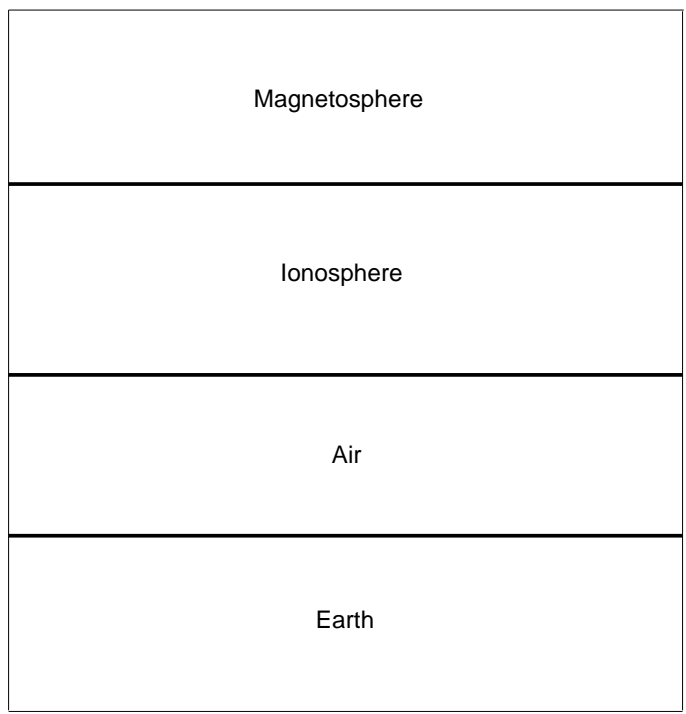

**Figure 2.** Schematic representation of the four geospheres.

Let us explain how two essentially different types of standing Alfvén waves can be excited in the ionosphere. Both types are characterized by a discrete spectrum. The difference is that in one case the frequencies of the spectral lines are proportional to the natural numbers 1, 2, 3, .., and in the other they are proportional to the odd numbers 1, 3, 5, ... .

Within the framework of ideal magnetohydrodynamics, the dispersion relation for Alfvén waves has the form

$$\omega = c_A k_{\parallel} \tag{1}$$

where $c_A = B / \sqrt{4\pi\rho}$, $k_{\parallel} = \mathbf{k}\mathbf{B} / B$ is the longitudinal component of the wave vector $\mathbf{k}$, $\rho$ is the plasma density [Alfvén, Falthammar, 1963]. Standing waves result from the reflection of waves from the upper and lower boundaries of the ionosphere. It is quite obvious that at the upper boundary of the layer there will be an antinode of oscillations of the electric field $\mathbf{E}$ and, accordingly, a node of oscillations of the magnetic field $\mathbf{b}$. As for the lower border, the situation is more complicated. The point is as follows.

Frequency (1) does not depend on the transverse component of the wave vector $\mathbf{k}_{\perp}$, i.e. we are dealing with a strongly degenerate spectrum. Generally speaking, the transverse structure of the wave field at a given frequency $\omega$ can be arbitrary. In reality, it depends on the conditions for the excitation of Alfvén waves.

Let us agree to call a standing wave thick if small values of the transverse component of the wave number prevail in the spectral expansion of the wave field in horizontal coordinates: $k_{\perp} h < 1$. Here $h$ is the thickness of the air gap, approximately equal to 90 km. With the opposite inequality ($k_{\perp} h > 1$), we will call the standing wave thin. It can be seen that the field of a thick standing wave penetrates to the Earth's surface. Here is the node of the field $\mathbf{E}$ and the antinode of the field $\mathbf{b}$. Knowing the location of nodes and antinodes at the upper and lower boundaries of the ionosphere, it is easy to make sure that the discrete frequencies of the IAR are proportional to the odd numbers of harmonics 1, 3, 5, and so on. If the wave is thin, then the air gap prevents the penetration of the field to the earth's surface. An antinode $\mathbf{E}$ and a node $\mathbf{b}$ will be located at the lower boundary of the ionosphere. Discrete oscillation frequencies of the ionosphere are proportional to the natural numbers 1, 2, 3, and so on.

Both versions of the spectral structure, in principle, can be observed in the ionosphere. However, odd harmonics are expected to be observed on the Earth's surface, since even-numbered harmonics do not penetrate the atmosphere.

At the end of this Section, we use Figure 2 to illustrate the nature of our simplifications. The two lower lines illustrate rather sharp boundaries between geospheres with significantly different electrodynamic properties, but the upper line has only a symbolic meaning. In reality, the ionosphere smoothly transforms into the magnetosphere. Nevertheless, the Alfvén wave in the Pc1 range, propagating from bottom to top, is reflected and propagates downward as a result of violation of the conditions of geometric optics applicability (see review [Potapov et al., 2021] and the literature cited therein). For simplicity, we have introduced a sharp interface, but in such a way that, in accordance with reality, an antinode of the electric field $\mathbf{E}$ of the standing Alfvén wave appears on it. Further, we consider the ionosphere and magnetosphere to be anisotropic but non-gyrotropic media. This is an admissible idealization everywhere, except for a thin layer in the lower part of the ionosphere. We will return to the question of the influence of the gyrotropy of the ionosphere lower edge in Section 4.

## 3. Data and analysis

So, we want to determine the frequency ratio of the IAR spectral lines observed on the earth's surface in order to test our hypothesis about the distribution of harmonic numbers over odd numbers. As a starting material, we used the dynamic spectra of the IAR, recorded at the mid-latitude observatory Mondy of ISTP SB RAS. The observation interval was taken from January 1 to February 28, 2017. Diurnal spectrograms were selected containing intervals of rather narrow spectral bands and a clearly pronounced first harmonic. A total of 12 days were chosen in this way. On each of the selected spectrograms, an hour mark was found, above which at least five harmonics were most clearly traced, including the first. The table shows the result of primary processing. The columns show the date, universal and local time of each measurement, as well as the measured frequencies of the first five harmonics.

**Table.** The measured frequencies (in hertz) of the first 5 harmonics on winter days 2017. The letters after the dates indicate the field component used for the measurements.

| Date | UT | LT | 1 | 2 | 3 | 4 | 5 |
|---|---|---|---|---|---|---|---|
| 02.01.2017-x | 9 | 16 | 0.4 | 1.1 | 1.75 | 2.5 | 3.3 |
| 04.01.2017-y | 10 | 17 | 0.55 | 1.4 | 2.25 | 3.2 | 4 |
| 10.01.2017-y | 11 | 18 | 0.5 | 1.95 | 3 | 4.22 | 5.4 |
| 11.01.2017-y | 10 | 17 | 0.5 | 1.45 | 2.25 | 3.25 | 4.2 |
| 13.01.2017-y | 9 | 16 | 0.45 | 1.1 | 1.95 | 2.5 | 3.3 |
| 21.01.2017-y | 11 | 18 | 0.6 | 1.7 | 2.75 | 3.9 | 5 |
| 22.01.2017-x | 12 | 19 | 0.6 | 1.7 | 2.85 | 4 | 5.1 |
| 23.01.2017-y | 9 | 16 | 0.3 | 0.9 | 1.5 | 2 | 2.7 |
| 29.01.2017-y | 11 | 18 | 0.55 | 1.55 | 2.5 | 3.5 | 4.5 |
| 30.01.2017-y | 10 | 17 | 0.35 | 1.15 | 1.9 | 2.6 | 3.45 |
| 10.02.2017-y | 12 | 19 | 0.65 | 1.8 | 2.8 | 3.9 | 5.05 |
| 22.01.2017-x | 14 | 21 | 0.55 | 1.9 | 2.9 | 3.95 | 5 |

In the process of further processing, we built triangular matrices for each of the events indicated in the table. The matrices were built according to the following rule. The first row indicated the frequencies of the higher harmonics in units of the frequency of the first harmonic; the second line indicated the frequencies of the higher harmonics in units of the second harmonic, and so on. For example, for the event 01/02/2017, the triangular matrix has the form

$$\begin{Vmatrix} 2.75 & 4.38 & 6.25 & 8.25 \\ 0 & 1.59 & 2.27 & 3.00 \\ 0 & 0 & 1.43 & 1.89 \\ 0 & 0 & 0 & 1.32 \end{Vmatrix} \qquad (2)$$

Finally, a summary matrix was calculated for all twelve events:

$$\begin{Vmatrix} 2.96 \pm 0.41 & 4.77 \pm 0.56 & 6.61 \pm 0.78 & 8.57 \pm 1.01 \\ 0 & 1.61 \pm 0.07 & 2.24 \pm 0.07 & 2.9 \pm 0.12 \\ 0 & 0 & 1.39 \pm 0.05 & 1.80 \pm 0.05 \\ 0 & 0 & 0 & 1.29 \pm 0.03 \end{Vmatrix} \qquad (3)$$

Average values and standard deviations are shown here.

## 4. Discussion and conclusion

We found an original method of upper triangular matrices for a uniform representation of the IAR spectral lines observed in the experiment. The idea of the method is based on the dependence of the matrix elements on the qualitative structure of the resonator, and their independence on the quantitative values of the spectral line frequencies.

Let us introduce a notation $T_{ij}$ for empirical matrices of the type (2), (3). Here the index $i$ is the ordinal number of the nonzero element of the matrix in the row with the number $j$. Both indices run through the values 1, 2, 3, ... Compare $T_{ij}$ with two model matrices

$$T_{ij}^1 = \frac{2i+1}{2j-1}, \quad T_{ij}^2 = \frac{i+1}{j}. \qquad (4)$$

Here the first (second) matrix refers to the idealized IAR model with an antinode (node) of the magnetic field $\mathbf{b}$ on the lower wall of the resonator. A close examination of the summary matrix (3) indicates that $T_{ij}$ is much closer to $T_{ij}^1$ than to $T_{ij}^2$. For a more definite judgment about this trend, we will use the following consideration.

An invariant characteristic of a matrix is its determinant $\det\left(T_{ij}\right)$. The determinant of a triangular matrix is equal to the product of the elements of its main diagonal [Gantmakher, 1988]. It is easy to see that

$$\det\left(T_{ij}^1\right) = \prod_{j=1}^{j_{max}} T_{1j}^1 = 2\,j_{max} + 1, \tag{5}$$

and

$$\det\left(T_{ij}^2\right) = \prod_{j=1}^{j_{max}} T_{1j}^2 = j_{max} + 1. \tag{6}$$

In the case of infinite matrices ($j_{max} \to \infty$), we have $\det\left(T_{ij}^1\right)/\det\left(T_{ij}^2\right) = 2$. In our case $j_{max} = 4$ and, thus, we have $\det\left(T_{ij}^1\right) = 9$ and $\det\left(T_{ij}^2\right) = 5$. Let us compare these numbers with the value $\det\left(T_{ij}\right) = 8.54$ obtained from the summary matrix (3). We see that $T_{ij}$ is much closer to $T_{ij}^1$ than to $T_{ij}^2$. For verification, we calculated the determinants separately for each of the twelve events shown in the table. The result was the following. The lower and upper sextiles are 7.7. and 9.2, respectively, and the median is 8.4. Thus, there is no doubt that, according to our measurements, the frequencies of the IAR spectral lines are distributed over odd harmonics 1, 3, 5, 7, 9.

We see some deviation of $\det\left(T_{ij}\right)$ from $\det\left(T_{ij}^1\right)$, but it is easy to explain. Let us use the analogy between Alfvén oscillations in an ionospheric resonator and string oscillations. If the string is homogeneous, then the oscillation frequencies satisfy simple numerical ratios. It is a different matter if the string is not uniform. In particular, in an inhomogeneous string, the oscillation harmonics become non-equidistant. Something similar is observed in real IAR, which explains the difference between $\det\left(T_{ij}\right)$ and $\det\left(T_{ij}^1\right)$.

We made our prediction based on the idealized IAR model and tested it experimentally. Within the framework of the same simplified model, it is expected that the spectral lines of the so-called *additional Alfvén resonator* [Dovbnya et al., 2013] will be distributed over a natural series of numbers. The same applies to the spectral lines of the IAR, if they are observed not on the earth's surface, but directly in the ionosphere using a satellite.

However, our idealized model needs correction, and this can lead to a possible change in predictions about the structure of the IAR spectrum. The point here is as follows: our model does not take into account the gyrotropic and dissipative properties of the lower ionosphere. Let us dwell briefly on this issue.

Generally speaking (but not in this particular case), gyrotropy is not necessarily accompanied by anisotropy or dissipation, these are the three independent properties of substance. But here an anisotropic absorbing layer of the ionosphere is situated almost at the gyrotropic layer. Thin enough layers of this kind are conveniently simulated by films. We can place our films at one and the same height (about 100 km) and use the Levi-Civita boundary condition, first introduced into electrodynamics in 1902 (e.g., see [Bateman, 1955; Guglielmi, Pokhotelov, 1996]). The beginning of research of the ULF electromagnetic phenomena in thin gyrotropic conducting films goes back to the monograph of Davydov [1971]. With regard to the IAR, the issue has been studied in many theoretical works (see, for example [Polyakov, Rapoport, 1981; Lysak, 1991; Guglielmi, Pohotelov, 1996; Pokhotelov et al., 2000; Demekhov et al., 2000; Lysak, Yoshikava, 2006; Fedorov et al., 2016]). Within the framework of this short article, we cannot delve into this delicate and complex problem, so we restrict ourselves here to the above references to the literature.

In conclusion, we want to express our confidence that experimental verification of hypotheses regarding the properties of IAR is necessary not only for the development of the theory, but also for replenishing knowledge about the structure and dynamics of the ionospheric plasma.

**Acknowledgements**. We express our sincere gratitude to T.N. Polyushkina and B. Tsegmed for their help in processing the material, to B.V. Dovbnya and N.A. Kurazhkovskaya for their interest in the work and fruitful discussions of the IAR problem. This work was carried out with financial support from the Russian Foundation for Basic Research in the framework of project No. 19-05-00574, as well as from IPE RAS and ISTP SB RAS state assignment programs of the Ministry of Science and Higher Education of Russia.